\documentclass[aps,showpacs,twocolumn,superscriptaddress]{revtex4}
\usepackage{epsf}
\usepackage{graphicx}
\usepackage{dcolumn}

\newcommand{\nk}{{\bf k}}
\newcommand{\np}{{\bf p}}
\newcommand{\nq}{{\bf q}}

\newcommand{\nP}{{\bf P}}

\begin{document}

\title{Induced Nucleon Polarization and Meson-Exchange 
Currents in $(e,e'p)$ Reactions }


\author{F. Kazemi Tabatabaei}

\affiliation{Departamento de F\'{\i}sica Moderna, 
          Universidad de Granada,
          Granada 18071, Spain}  
\author{ J.E. Amaro}
\affiliation{Departamento de F\'{\i}sica Moderna, 
          Universidad de Granada,
          Granada 18071, Spain}
\author{J.A. Caballero}
\affiliation{Departamento de F\'\i sica At\'omica, Molecular y Nuclear 
          Universidad de Sevilla, Apdo. 1065, 
          Sevilla 41080, Spain }

\begin{abstract}

Nucleon recoil polarization observables in $(e,e'\vec{p})$ reactions 
are investigated using a semi-relativistic 
distorted-wave model which includes one- and
two-body currents with relativistic corrections.
Results for the induced polarization asymmetry are 
shown for closed-shell nuclei and a comparison with 
available experimental data for $^{12}$C is provided.
A careful analysis of meson exchange currents shows that they
may affect significantly the induced polarization for high missing momentum.
\end{abstract}



\maketitle


\section{Introduction}


Measurements of the polarization of the ejected proton in
$(e,e'\vec{p})$ reactions~\cite{Woo98} provide valuable
information on the nucleus complementary to that extracted from
unpolarized experiments~\cite{Fru84,Bof93,Kel96,Bof96}. In fact
a new set of 8 spin-dependent response functions that
present different sensitivities to the various ingredients of the 
reaction mechanism enter in the general analysis~\cite{Pic87,Pic89,Ras89}.
A richer source of information on nucleon
properties inside the nucleus is thus embedded into the spin-dependent
nuclear responses.

In a previous work~\cite{Kaz03} we have developed a model aiming to
provide a systematic investigation of spin-dependent observables in
$(\vec{e},e'\vec{p})$ reactions.  Relying on the distorted wave
impulse approximation (DWIA), our approach includes in addition
two-body meson exchange currents (MEC) and relativistic corrections
based on the semi-relativistic form of the electromagnetic currents
derived in the last years~\cite{Ama96a,Ama98a,Ama02c,Ama03}.

In Ref.~\cite{Kaz03} the full set of polarized response functions was
computed and analyzed for intermediate to high values of the momentum
transfer, $q$, at the quasielastic peak. Their dependence on the model
of final state interactions (FSI) was studied and the effects of MEC
were evaluated. The emphasis was placed on the proton polarization
induced by polarized electrons, i.e., on the transferred polarization
asymmetries $P'_{l,s}$, which only contribute when the initial
electron beam polarization is measured. These transferred asymmetries
survive in the plane wave impulse approximation (PWIA) limit and may
provide ideal tools for studying the electromagnetic nucleon form
factors in the nuclear medium~\cite{Mal00,Ryc99,Kel99,Cris03}.

The focus of this paper is the analysis of the properties displayed by
the polarization observables induced by {\em unpolarized} electrons,
i.e., the induced polarization asymmetry $\nP$ which, contrary to the
transferred asymmetries, is zero in PWIA.  In fact, since the target
nucleus is unpolarized the electron can hit with equal probability
nucleons with all spin orientations along their orbits. In absence of
FSI, these nucleons leave the nucleus as plane waves with the same
amplitude, hence giving no net induced polarization. The situation
clearly differs when FSI are considered in the description of the
process; first, because of the relation between the spin direction and
the nucleon location in the orbit which implies different FSI
strengths for different spin orientations due to the central part of
the optical potential (mainly the imaginary absorptive term), and
second, because of the explicit spin dependence of the spin-orbit
interaction in the optical potential.

One of the goals of this work is to evaluate the impact of the
two-body MEC over the induced polarization components and hence to
analyze the validity of the impulse approximation (IA), i.e., one-body
currents only.  We are guided by previous studies~\cite{Ama03b,Kaz03}
where the role of MEC on asymmetry observables has been found to be in
general small for low missing momentum $p$. This result is in part due
to the occurrence of an effective cancellation of MEC effects between
the numerators and denominators involved in the polarization
ratios. The same applies to FSI effects.  However for higher values of
the missing momentum the effective cancellation does not occur and MEC
(likewise FSI) effects can be important. Other ingredients not
included in the present model, such as correlations~\cite{Maz02} and
relativistic nuclear dynamics~\cite{Cris03,Udi00}, have been also
shown to sizeably affect the polarization asymmetries at high $p$.
 
The structure of the paper is as follows: in Sec. II we shortly
present our distorted wave model. In Sec. III we discuss the results obtained
for the induced polarization asymmetry and compare with available data. 
Finally, the conclusions are drawn in Sec. IV.


\section{DWIA Model for $(e,e'\vec{p})$}


\begin{figure}[tb]
\begin{center}
\leavevmode
\def\epsfsize#1#2{0.9#1}
\epsfbox[190 480 430 700]{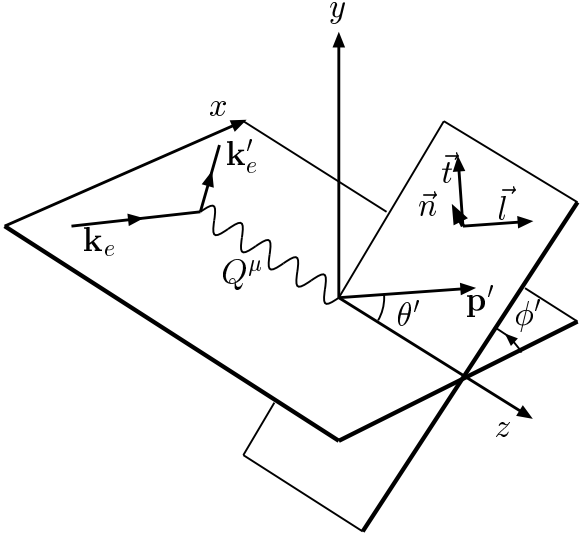}
\end{center}
\caption{Coordinate system used in the $(e,e'\vec{p})$ reaction. The
$x$-$z$ coordinates span the scattering plane, with the $z$-axis
pointing along the momentum transfer $\nq$. The proton polarization is
described in the barycentric system $(l,t,n)$, where the ejected
nucleon momentum, $\np'$, expands the $\vec{l}$ direction, while the
$\vec{l}$ and $\vec{t}$ vectors expand the reaction plane.  Finally
the normal vector $\vec{n}$ is defined by $\nq\times\np'$.  }
\end{figure}

We refer to our previous works~\cite{Kaz03,Ama03b} and references
therein for details on the model. Here we just set up some general
definitions of interest for the reader and for the discussion that
follows.  In the $(e,e'p)$ process
sketched in Fig.~1, we consider an electron with four-momentum
$K_e^\mu=(\varepsilon_e,\nk_e)$ that scatters off a nucleus
transferring a four-momentum $Q^{\mu}=(\omega,\nq)$. The electron
scattering angle is $\theta_e$.  A proton with momentum $\np'$ and
exit solid angle $\Omega'=(\theta',\phi')$ is detected in coincidence
with the outgoing electron. The proton spin polarization along an
arbitrary, unitary vector $\vec{s}$ is also measured. We assume that
the residual nucleus is left in a discrete state, and neglect the
recoil.  The cross section for this process can be written in the Born
approximation and extreme relativistic limit for the electron as
\begin{eqnarray}
\Sigma
&\equiv&\frac{d\sigma}{d\epsilon'_e d\Omega'_e d\Omega'}
\\
&=&
K\sigma_M\left(v_LR^L+v_TR^T+v_{TL}R^{TL}+v_{TT}R^{TT}\right)
\nonumber
\end{eqnarray}
where $\sigma_M$ is the Mott cross section, $K$ is the kinematic
factor $m_Np'/(2\pi\hbar)^3$ (being $m_N$ the nucleon mass) and the
$v_\alpha$ coefficients, $\alpha=L,T,TL,TT$ are the usual ones arising
from the leptonic tensor \cite{Ras89}. Finally, the exclusive response
functions $R^{\alpha}$ are linear combinations of the Hadronic tensor
and hence they contain all the pertinent information on the nuclear
reaction mechanism.

In the present paper we make use of the semi-relativistic distorted
wave model developed in Refs.~\cite{Kaz03,Ama03b} whose basic
ingredients are the following: i) the final proton state is described
by a solution of the Schrodinger equation with a non-relativistic
optical potential, but assuming the relativistic energy-momentum
relation, thus we effectively solve a Klein-Gordon kind
equation. ii) Semi-relativistic (SR) operators are used for the
one-body (OB) electromagnetic current and two-body MEC.  These have
been obtained by expanding the corresponding relativistic operators to
first order in the missing momentum over the nucleon mass $p/m_N$
(being $\np=\np'-\nq$), while the exact dependence on $(\omega,\nq)$
is maintained~\cite{Ama02c,Ama03}.  We consider the one-pion exchange
diagrams of seagull (S or contact), pion-in-flight (P or pionic) and
$\Delta$-isobar kinds.

\begin{figure}[tb]
\begin{center}
\leavevmode
\def\epsfsize#1#2{0.6#1}
\epsfbox[85 350 475 800]{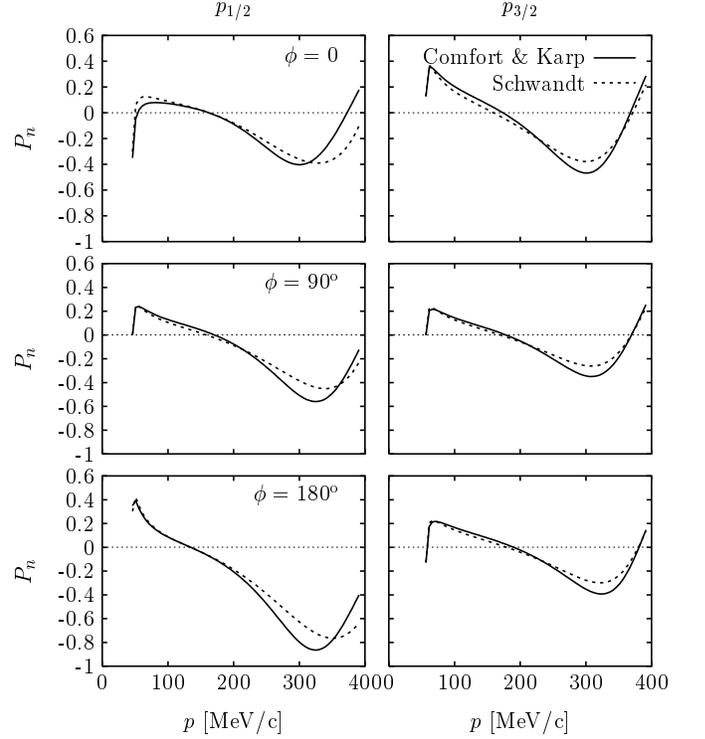}
\end{center}
\caption{ Proton induced polarization in the $n$-direction for
knock-out from the $p$ shells of $^{16}$O. The kinematics are $q=460$
MeV/c, $\omega=100$ MeV, $\theta_e=30^{\rm o}$. The dependence on FSI
is analyzed by comparison of two-optical potentials. }
\end{figure}

\begin{figure}[tb]
\begin{center}
\leavevmode
\def\epsfsize#1#2{0.6#1}
\epsfbox[55 500 455 800]{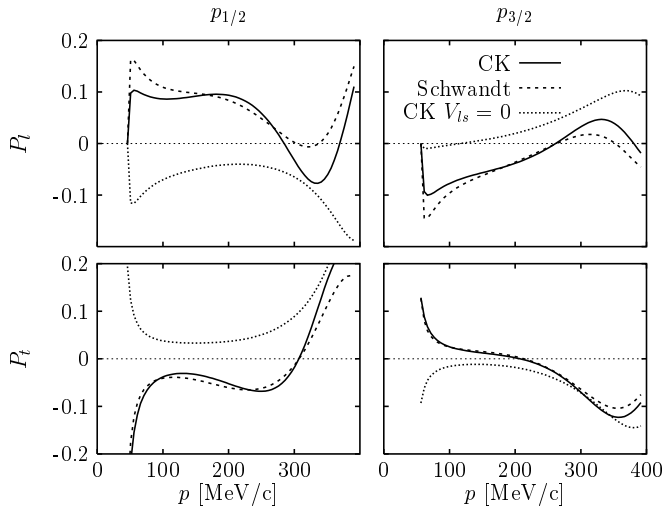}
\end{center}
\caption{ The same as Fig.~2 for the $P_l$ and $P_t$ polarization
components and for $\phi=90^{\rm o}$. 
 Note that these observables are zero for in-plane
emission (i.e., $\phi=0$ or $180^{\rm o}$). Dotted lines correspond to the 
Comfort-Karp potential but neglecting its spin-orbit dependence. }
\end{figure}

The induced polarization asymmetry $\nP$, which is the focus of this paper, is
defined by
\begin{equation}
\Sigma =\frac12
\Sigma_{unpol}\left(1+\nP\cdot\vec{s}\right) \, .
\end{equation}
The vector $\nP=(P_l,P_t,P_n)$ is usually set in the barycentric
coordinate system, referred to the reaction plane, as longitudinal
($\vec{l}$), transverse or sideways ($\vec{t}$), and normal
($\vec{n}$) directions defined in  Fig. 1.

These induced polarization components can be written in the form
\begin{eqnarray}
P_n
&=&
\frac{2}{W_0}
\left(v_LW^L_n+v_TW^T_n \right.
\nonumber\\
&&\mbox{}
\left.+v_{TL}\cos\phi W^{TL}_n +v_{TT}\cos2\phi W^{TT}_n\right) \, ,
\label{Pn}
\\
P_l
&=&
\frac{2}{W_0}
\left(v_{TL}\sin\phi W^{TL}_l +v_{TT}\sin2\phi W^{TT}_l\right) \, ,
\label{Pl}
\\
P_t
&=&
\frac{2}{W_0}
\left(v_{TL}\sin\phi W^{TL}_t +v_{TT}\sin2\phi W^{TT}_t\right) \, ,
\label{Pt}
\end{eqnarray}
where $\phi=\phi'$ is the azimuthal angle of $\np$, and 
we have defined the function
\begin{equation}
W_{0}\equiv
      v_LW^L_{0}+v_TW^T_{0}
       +v_{TL}\cos\phi W^{TL}_{0} +v_{TT}\cos2\phi W^{TT}_{0},
\end{equation}
and the unpolarized $W^\alpha_{0}$ and polarized 
$W^\alpha_i$ reduced response functions have been introduced.
The role of the various ingredients of our
model (FSI and MEC) over the separate response functions was analyzed
in \cite{Ama03b,Kaz03}. In the following we show results for the 
induced polarization components for selected kinematical conditions.

\section{Results for the induced polarization}

Since the induced polarization is zero in absence of FSI, this
observable is expected {\em a priori} to be specially sensitive to
details of the optical potential used to describe the final proton
state. Results of FSI model dependences are presented in Figs.~2 and 3
for proton knock-out from the $p_{1/2}$ and $p_{3/2}$ shells in
$^{16}$O. Quasi-perpendicular kinematics is
considered~\cite{Chi91,Spa93}, with $q=460$ MeV/c and $\omega=100$
MeV, corresponding closely to the quasi-elastic peak.  The electron
scattering angle is $\theta_e=30^{\rm o}$. In Fig.~2 we show results
for the normal component $P_n$ for three values of the proton
azimuthal angle $\phi=0$, $90^{\rm o}$ and $180^{\rm o}$ (see
definition in Fig.~1). Two optical potentials widely used in the
literature to describe these reactions are considered: the
Schwandt~\cite{Sch82} potential with dashed lines and the Comfort \&
Karp~\cite{Com80} one with solid lines.  Note that the Schwandt
potential has been extrapolated here to $A=16$, since it was fitted
for heavier nuclei.  These two potentials differ in their spin-orbit
dependence. Whereas the Comfort-Karp potential includes a purely real
term, Schwandt's has also an imaginary part; moreover the real part of
the Comfort potential is more attractive near the nuclear
surface. Concerning the dependence on the real and imaginary parts of
the central potential, Schwandt's is more attractive and has less
absorption. However, in their gross features the two potentials do not
present remarkable discrepancies. This explains why at low missing
momentum they provide similar predictions for $P_n$, starting to
differ for higher $p$-values $p>200$ MeV/c.

\begin{figure}[tb]
\begin{center}
\leavevmode
\def\epsfsize#1#2{0.6#1}
\epsfbox[55 350 455 800]{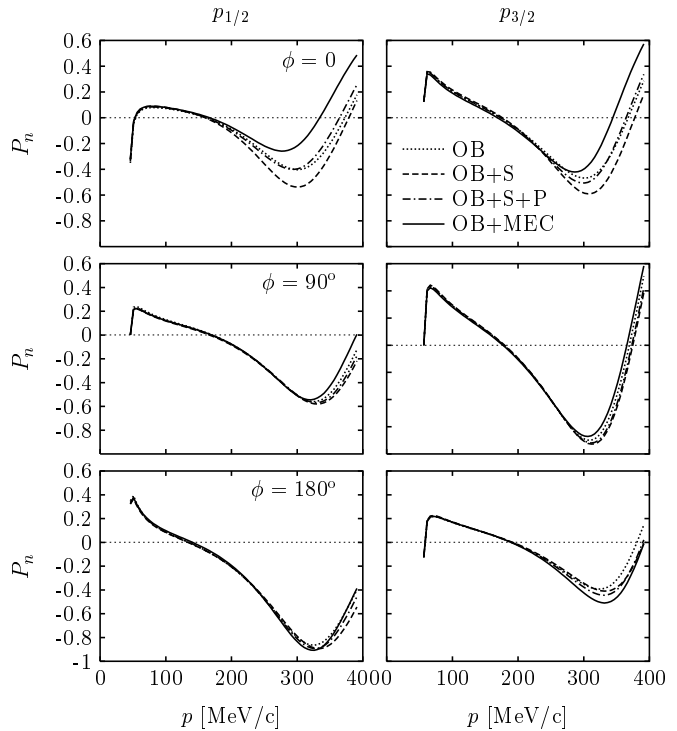}
\end{center}
\caption{MEC effects over the  
proton induced polarization in the $n$-direction for
knock-out from the $p$ shells of $^{16}$O,  $q=460$ MeV/c,
$\omega=100$ MeV, and $\theta_e=30^{\rm o}$.}
\end{figure}

The corresponding $P_l$ and $P_t$ polarization components are shown in
Fig.~3. Note from Eqs.~(\ref{Pl},\ref{Pt}) that these components are
zero for in-plane emission ($\phi=0,180^{\rm o})$, hence we only
present results for out-of-plane kinematics, $\phi=90^{\rm o}$. The
biggest differences between both potentials show up in $P_l$ (upper
panels) for low and high values of the missing momentum, while $P_t$
(lower panels) exhibits less dependence to details of the potential.
The reason why $P_l$ is more sensitive to the details is not clear.
However both polarizations are crucially dependent on the interaction,
since they are strictly zero in PWIA.  This is clearly illustrated in
Fig.~3 where we show with dotted lines the results obtained with the
Comfort-Karp potential but neglecting its spin-orbit dependence. The
drastic change produced in the two polarizations shows that both
observables depend equally on the global form of the interaction.
However, at the kinematics selected, $P_l$ presents a slightly
stronger sensitivity to the ``fine'' details of the potential

In the following we analyze MEC effects restricting
ourselves to the use of the Comfort \& Karp potential.  Discussion of
results for the Schwandt potential follows similar trends.

\begin{figure}[tb]
\begin{center}
\leavevmode
\def\epsfsize#1#2{0.6#1}
\epsfbox[60 500 450 800]{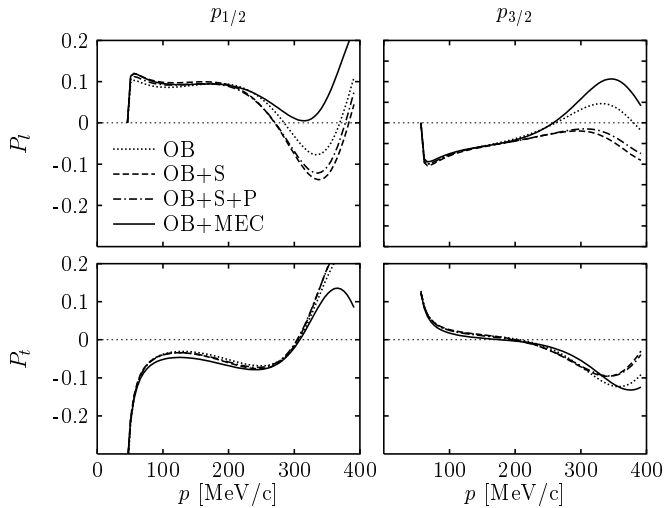}
\end{center}
\caption{ 
The same as Fig.~4 for the $P_l$ and $P_t$ polarization
components and for $\phi=90^{\rm o}$. 
}
\end{figure}

In Figs.~4--7 we present the impact of MEC upon the induced
polarization components for two values of the momentum transfer.
Fig.~4 displays the $P_n$ polarization for intermediate $q=460$ MeV/c
and $\omega=100$ MeV.
In addition to the OB calculation (dotted lines), in each panel of the
figure we show three more curves corresponding to the additional
contribution of the several MEC: seagull (OB+S), seagull plus pionic
(OB+S+P) and total MEC (OB+S+P+$\Delta$).  As shown, for low missing
momentum MEC contributions are in general negligible, and tend to
increase as $p$ goes higher. In particular, for $\phi=0$ (top panels)
MEC are shown to modify significantly the results of $P_n$ for $p>200$
MeV/c. This effect being larger for the $p_{1/2}$ shell.  For the
other $\phi$-values selected, 
 MEC are smaller for all missing
momenta. This outcome, which is specific of the kinematics selected,
can be ascribed to a cancellation of the two-body effects in
$P_n$. Note that the contribution of the interference $TL$ response
for $\phi=180^{\rm o}$ is just opposite to that one occurring at
$\phi=0$, while for $\phi=90^{\rm o}$ the $TL$ 
response does not enter, see Eq.~(\ref{Pn}).

Larger MEC effects for high missing momentum, $p>200$ MeV/c, are shown
in Fig.~5 in the case of the longitudinal and transverse induced
polarizations. Note that $P_l$ (top panels)
may even change its global sign when MEC are considered.  As in the
previous case of $P_n$, the role of MEC for $p<200$ MeV/c is again
negligible. From these results, we may conclude that the description
of the induced polarization observables within the IA, i.e., with OB
current operators only, is expected to be quite acceptable in this
low-$p$ regime.

Concerning the separate role played by the S, P and $\Delta$-currents,
results in Figs.~4 and 5 for high missing momentum, $p>200$ MeV/c,
show that the seagull contribution is opposite to those provided by
the pionic and $\Delta$ currents.  Quantitatively the importance of
the three currents upon $P_n$ for $\phi=0$ is similar. Note also that
the P and S contributions tend to cancel in this case.  On the
contrary, for the transverse induced polarizations, particularly for
$P_l$, the $\Delta$ current is clearly dominant, being almost
negligible the pionic one.

\begin{figure}[tb]
\begin{center}
\leavevmode
\def\epsfsize#1#2{0.6#1}
\epsfbox[80 360 470 800]{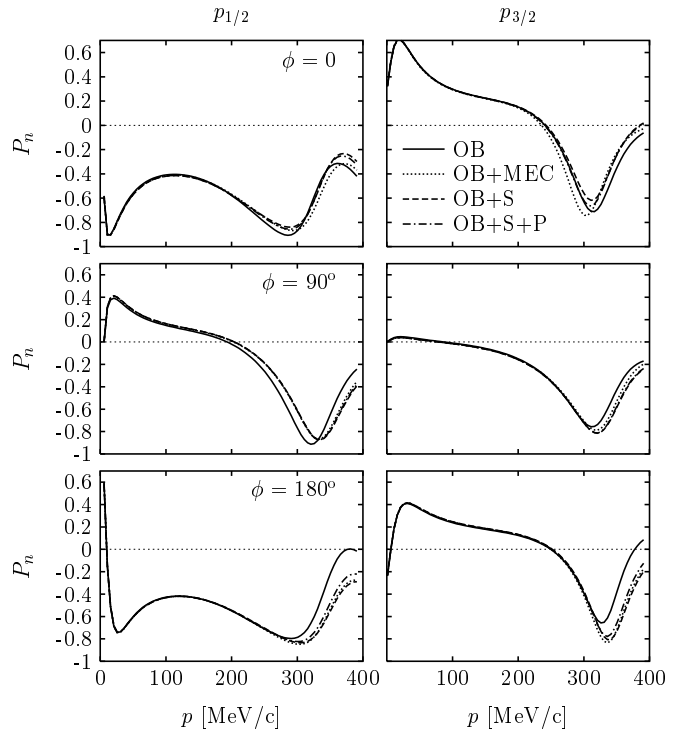}
\end{center}
\caption{MEC effects over the  
proton induced polarization in the $n$-direction for
knock-out from the $p$ shells of $^{16}$O,  $q=1$ GeV/c,
$\omega=450$ MeV, and $\theta_e=30^{\rm o}$.}
\end{figure}

Results for higher values of $q=1$ GeV/c and $\omega=450$ MeV are
shown in Figs.~6--7. This kinematics corresponds to a recent
experiment at TJLab~\cite{Gao00}, where the $A_{TL}$ asymmetry
and the transfer polarization were measured.
The use of the
present semi-relativistic model for this kinematics is justified by a
comparison with the relativistic distorted wave impulse approximation
(RDWIA) calculation of Udias {\em et al.}~\cite{Cris03,Udi99,Udi01}.
Although the dynamical relativistic effects
that occur in RDWIA have been shown to be important in
the high missing momentum region, 
the results in Figs.~6 and 7 constitute an
indication of what kind of effects can be expected from MEC in this
region.

\begin{figure}[tb]
\begin{center}
\leavevmode
\def\epsfsize#1#2{0.6#1}
\epsfbox[85 500 475 800]{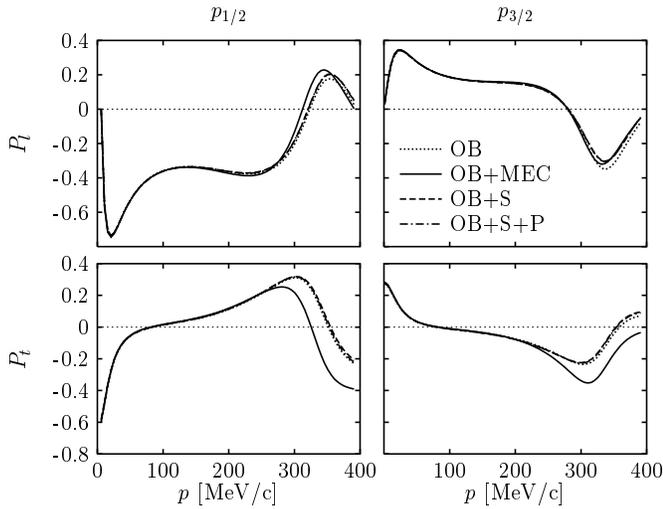}
\end{center}
\caption{ 
The same as Fig.~6 for the $P_l$ and $P_t$ polarization
components, for $\phi=90^{\rm o}$.
}
\end{figure}

As in the previous kinematics, MEC contributions are negligible for
low missing momentum, $p\leq 200$ MeV/c. This strong MEC suppression
is in part due to the behavior of the pion-nucleon form factor at high
$|Q^2|$. For higher $p$-values, $p > 300$ MeV/c, MEC start to be
important giving a significant contribution for $P_n$ at
$\phi=180^{\rm o}$ and $P_t$ at $\phi=90^{\rm o}$.  Note the
difference with the previous kinematics where the largest MEC effects
were exhibited by $P_n$ at $\phi=0$ and $P_l$ ($\phi=90^{\rm o}$).
Particularly noteworthy is also the clear dominance of the $\Delta$
current over the P and S terms.  These MEC effects are similar to the
ones found over the $A_{TL}$ asymmetry for the same
kinematics~\cite{Ama03b}.

\begin{figure}[t]
\begin{center}
\leavevmode
\def\epsfsize#1#2{0.7#1}
\epsfbox[130 390 430 780]{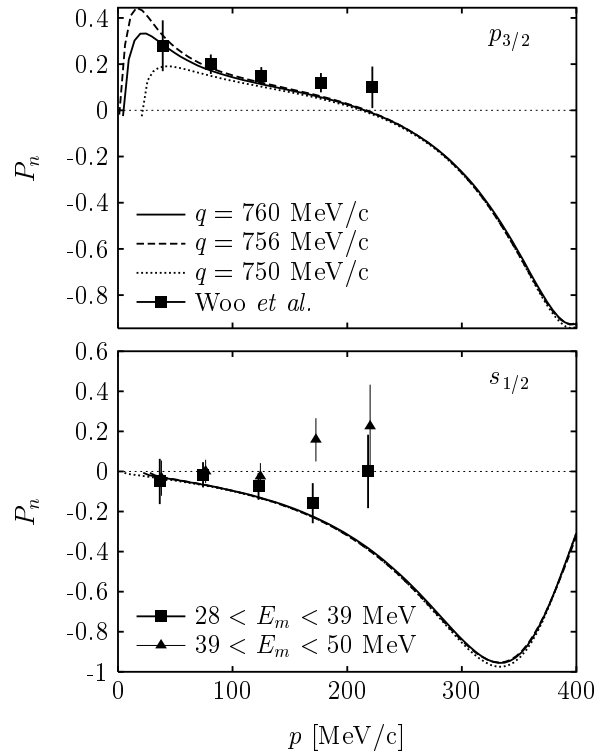}
\end{center}
\caption{ Normal induced polarization for $^{12}$C computed in DWIA
with OB current operators only. The electron energy is
$\epsilon_e=579$ MeV, and $\phi=0^{\rm o}$ .  Three kinematics close
to the quasi-elastic peak are displayed: $(q,\omega)=$ (760 MeV/c, 290
MeV), (756 MeV/c, 284 MeV), and (750 MeV/c, 294 MeV).  The
experimental data, corresponding to $|Q^2|=0.5$ (GeV/c)$^2$, are from
Ref.  \protect\cite{Woo98}.}
\end{figure}

In Fig.~8 we show results for the normal polarization of proton
knock-out from the two shells in $^{12}$C, including only OB
electromagnetic operators.  We have chosen three sets of kinematics
following closely those of Ref.~\cite{Woo98}, $q\simeq 760$ MeV/c and
$\omega \simeq 290$ MeV.  These values correspond nearly to the
quasielastic peak.  Since several sets of values have been used in the
literature when comparing with the corresponding experimental data, in
Fig.~8 we show three curves for three slightly diverse $(q,\omega)$
sets. The results for the $p_{3/2}$ shell (upper panel) illustrates
that extreme caution is needed before final conclusions can be
drawn. In fact, the experimental point at $p\simeq 40$ MeV/c is
extremely sensitive to the missing momentum region allowed by the
kinematics.  For exact quasi-elastic conditions, this region begins at
$p=0$, corresponding to forward emission of a nucleon with momentum
$p'=q$.  A slight change of $(q,\omega)$ can shift the allowed region
by more than $25$ MeV/c.  The large error bars in the first data point
are then reminiscent of the large instability of $P_n$ under tiny
kinematical variations. The remaining data points located in the
region of larger missing momentum, and the case of the $s_{1/2}$
shell, where a great stability exists, are of more physical interest
for the analysis of two-body currents.

\begin{figure}[tb]
\begin{center}
\leavevmode
\def\epsfsize#1#2{0.7#1}
\epsfbox[140 390 430 780]{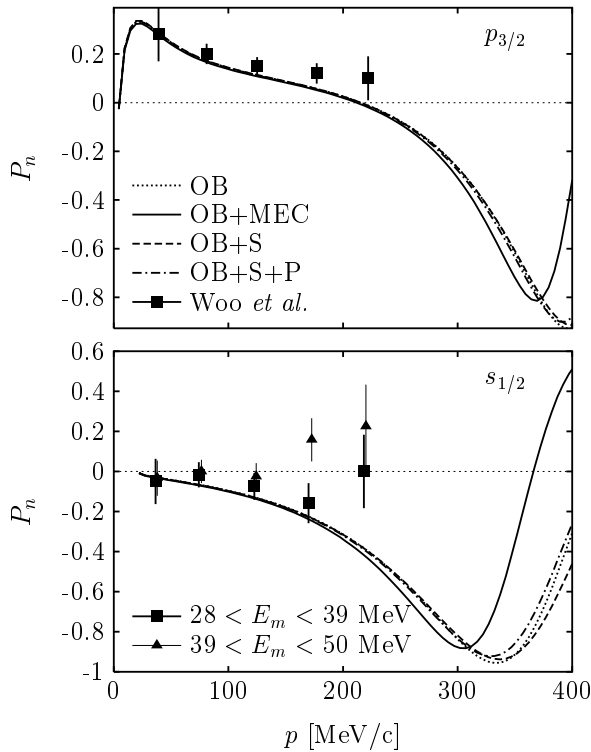}
\end{center}
\caption{ MEC effects over the normal induced polarization for
$^{12}$C. The kinematics correspond to Fig. 8 with 
$(q,\omega)=$ (760 MeV/c, 290 MeV).  
The experimental data are from Ref.  \protect\cite{Woo98}.  }
\end{figure}

Regarding MEC effects upon $P_n$ in $^{12}$C, we show in Fig.~9 the
comparison between our calculations including the OB and the several
pieces of the two-body current. In the region of low missing momentum,
$p<200$ MeV/c, where experimental data are located, MEC contributions
are negligible for both shells. As in the case of $^{16}$O, here MEC
lead to significant effects, particularly due to $\Delta$ which gives
the main contribution, in the high missing momentum region $p > 300$
MeV/c.  Results in Fig.~9 also show that the SR distorted wave model
calculations agree nicely with data. Moreover, the discrepancy with
results obtained within the RDWIA framework \cite{Udi00,Joh99}, 
which is better suited to
describe these high $(q,\omega)$ data, is small
in the low missing momentum region, and begin to be important for
$p \sim  200$ MeV/c.
For larger missing
momenta, $p>300$ MeV/c, dynamical relativity, not included in our
calculations, plays an important role and differences between RDWIA
and SR-DWIA calculations increase. Note however, that MEC make also a
very significant effect in this high-$p$ region.

To finish we compare our results with previous calculations of MEC
effects over the induced polarization asymmetries, namely with the
models developed by the Gent~\cite{Ryc99} and Pavia~\cite{Bof90}
groups. In the Gent calculation~\cite{Ryc99} MEC contributions upon
$P_n$ for $^{12}$C were also found to be very small for low missing
momentum, while they increase importantly, particularly due to
$\Delta$, for high $p$.  Discrepancies with our results emerge because
of the different models used to describe FSI; the Gent group makes use
of a real potential without absorption.  In the case of the Pavia
calculation~\cite{Bof90}, the induced polarization was evaluated for
low missing momentum, $p<200$ MeV/c, and FSI were computed by means of
complex phenomenological optical potentials.  Their $P_n$ results for
$^{16}$O with the Giannini \& Ricco optical potential~\cite{Gia76},
and a kinematics close to the one of Fig.~4, are similar to ours. MEC
effects were found to be small in general, specially for the $p_{3/2}$
shell, while for $p_{1/2}$ some visible differences, strongly
dependent on the specific optical potential used, show up even within
this low-$p$ regime. This outcome contrasts with our calculations
which do not show any specific difference between both shells.


\section{Conclusions}


In this work we have analyzed the induced polarization asymmetry of
knocked-out protons in exclusive $(e,e'p)$ reactions to discrete
residual nucleus states. We have used a semi-relativistic distorted wave model
including relativistic corrections to the one- and two-body currents
as well as relativistic kinematics. We have applied the model to
proton knock-out from the outer shells of $^{16}$O and have compared with
the experimental data available for $^{12}$C.

Regarding FSI, the $P_n$ polarization is
little dependent on the details of the optical potential for low
missing momentum, while its dependence increases for high $p$.  The
longitudinal $P_l$ polarization is more sensitive to details of the
potential both for low and high values of $p$, while the sideways $P_t$
polarization does only show tiny FSI uncertainties for $p>300$ MeV/c.

Concerning MEC, they are negligible for low $p<200$ MeV/c, but they
can importantly change the induced polarization components for high
$p$. In this regime the largest MEC contributions are found for
intermediate values of the momentum transfer. The effects upon $P_n$
are in accordance with the Gent calculation, even if differences
emerge due to the different FSI models used. The comparison with the
older Pavia calculation, which gives rise to some peculiar differences
between the two $p$-shells in $^{16}$O, is more troublesome.

The present calculation justifies the validity of the impulse
approximation for $p<200$ MeV/c, while it emphasizes the fact that,
besides dynamical relativity, other
effects beyond the IA as MEC
are also expected to contribute sizeably for high missing momentum. 
New experimental data for these observables in this
regime would be welcomed to explore this physics.

\section*{Acknowledgments}
This work was partially supported by funds provided by DGI (Spain) and
FEDER funds, under Contracts Nos BFM2002-03218, BFM2002-03315 and
FPA2002-04181-C04-04 and by the Junta de Andaluc\'{\i}a.


\end{document}